\documentclass[11pt]{article}
\usepackage[T1]{fontenc}
\usepackage{libertinus}
\usepackage{libertinust1math}

\usepackage[latin,english]{babel}
\usepackage[hang,flushmargin]{footmisc}
\usepackage{xurl}
\usepackage{tabularx}
\usepackage[pdftex,bookmarks,colorlinks=true,citecolor=blue,linkcolor=blue,urlcolor=blue]{hyperref}
\usepackage{enumitem}
\usepackage{ellipsis}
\usepackage{ragged2e}

\newenvironment{ul}{\begin{itemize}[nosep,leftmargin=*]}{\end{itemize}}

\newenvironment{ulv}{\begin{itemize}[nosep,itemsep=3pt,leftmargin=*]}{\end{itemize}}

\title{Measuring social consensus}
\author{David Flater\\
  National Institute of Standards and Technology, U.S.A.\\
  david.flater@nist.gov}
\date{2024-11-18}

\begin{document}
\selectlanguage{english}
\pagestyle{plain}
\maketitle
\footnotetext{Official contribution of the National Institute of Standards
  and Technology (NIST); not subject to copyright in the United States.  The
  opinions, recommendations, findings, and conclusions in this publication do
  not necessarily reflect the views or policies of NIST or the United States
  Government.}

\renewcommand{\sectionautorefname}{Section}
\renewcommand{\subsectionautorefname}{Section}
\renewcommand{\subsubsectionautorefname}{Section}
\renewcommand{\paragraphautorefname}{Section}

\newcolumntype{G}{>{\raggedright\arraybackslash}X}

\begin{abstract}
\noindent
Many organizations describe their processes as consensus-driven, but there is
no consensus on the definition of consensus.  Qualitative definitions of
consensus prioritize social phenomena like ``unity'' that are not necessarily
measurable.  Quantitative definitions of consensus derive from numbers of
votes and can be realized in software.  When unity and cooperation become
unobtainable for any reason, measuring consensus as a quantity (an amount of
agreement) is a reasonable adaptation to alleviate gridlock and possibly
avoid escalation of conflicts.  This article investigates the metrology of
social consensus.
\end{abstract}

\section{Introduction}

\begin{quote}
  The first principle of republicanism is that the lex majoris partis is the
  fundamental law of every society of individuals of equal rights:  to
  consider the will of the society enounced by the majority of a single vote
  as sacred as if unanimous, is the first of all lessons in importance, yet
  the last which is thoroughly learnt.  This law once disregarded, no other
  remains but that of force, which ends necessarily in military
  despotism.  \cite{Jefferson_1878}
\end{quote}

Consensus is a Latin word.  Its primary definition in the
\foreignlanguage{latin}{Thesaurus linguae Latinae} is
``\foreignlanguage{latin}{concordia, convenientia, assensus}''
\cite[defn.~I]{TLL}---concord, agreement, assent.  In etymology of the
English word, the Oxford English Dictionary describes the Latin word as
``agreement, accord, unanimity, concord, harmony, sympathy''
\cite{OED3-consensus}.

In modern practice it has come to mean different things.  Some forms of
consensus can be assessed only qualitatively, such as the overall sentiment
that a group leader must articulate in order to maintain the group's
cohesion.  Other forms of consensus can be measured by voting, albeit with
limitations that are a characteristic of all voting methods.

When questions become politicized and cooperation breaks down, qualitative
assessments of consensus can face direct challenge from those who feel they
are on the losing end of a decision.  What does one do in these situations?
War, whether political, procedural, legal, or total, is always a threat.  But
one action that may avert war is to \emph{measure} the consensus so that
those on the losing end are faced with empirical evidence that they have
little support.

The path to reliable measurements begins with qualitative concepts like
\emph{hot} and \emph{cold}.  Through a process of refinement and compromise,
eventually qualitative concepts are replaced by a quantity (thermodynamic
temperature in this example).  A measurement method is then devised that,
through a similar process of refinement and compromise, is made repeatable,
reproducible, and objective.  Fitness for purpose is not guaranteed; e.g.,
setting a particular air temperature does not necessarily achieve the goal of
making a person feel neither hot nor cold.  But if the resulting quantity and
measurement method serve the purposes of the original qualitative concepts,
they can then replace them and, thus, remove subjectivity from an assessment.

Qualitative and quantitative forms of consensus fit into this story.
Consensus as a quantity would be an amount of agreement, for which voting is
a measurement method.  The goal is then to refine consensus to something that
can be evaluated in a repeatable and reproducible process that can be
automated, factoring out the qualitative, social, and emotional aspects that
are subjective—but also to remain aware of the compromises made and fitness
for the original purposes.

This article investigates the nature and limitations of the measurement model
of consensus.  The remainder proceeds as follows.
\autoref{newsec2} looks at qualitative definitions of consensus and
characterization methods.
\autoref{newsec3} proceeds to quantitative definitions and methods.
\autoref{newsec4} describes different measurement scenarios and process concerns.
\autoref{newsec5} addresses uncertainty for those measurements.
\autoref{newother} briefly describes other uses of consensus.
Finally, \autoref{newsec6} gives conclusions.

\section{Qualitative consensus}
\label{newsec2}

\subsection{Definitions}

Qualitative definitions of consensus are those that cannot be realized by an
algorithm.  They prioritize social phenomena like ``unity'' or the mood of
the crowd that are not necessarily measurable by counting votes and might
even be undermined by the process of voting.

The phrase ``unity, not unanimity'' appears in many contexts where group
cohesion is essential to effective leadership.  Nevertheless, it remains the
case that one active dissenter is all it takes to \emph{block} agreement—the
same as if an explicit unanimous vote were required.

Randy Schutt writes:  ``Consensus is a process for deciding what is best for
a \textbf{group}.  The final decision is often not the first preference of
any individual in the group, and many may not even like the final result.
But it is a decision to which they all \textbf{consent} because they know it
is the best one for the group.''  This form of consensus ``is a process for
people who want to work together honestly in good faith to find good
solutions for the group.  It cannot be used by people who do not, can not, or
will not cooperate'' \cite{2007_Schutt_ConsensusNotUnanimity}.

The Common Wheel Collective explains:

\begin{quote}
Consensus is not just the end result of the group's decision-making process,
or the part where a vote is taken and the vote is unanimous, barring any
blocks or stand-asides.  The consensus process has to be built into the
entire structure of the group or organization and form the basis for all of
its activities and basic operation.  This is true for all egalitarian
collectives, even those who accept some form of majority vote in their
decision-making and may therefore not strictly be defined as operating by
consensus.  \cite{CWC}
\end{quote}

The Internet Engineering Task Force (IETF) guidelines and procedures state,
``IETF consensus does not require that all participants agree although this
is, of course, preferred.  In general, the dominant view of the working group
shall prevail.  (However, it must be noted that `dominance' is not to be
determined on the basis of volume or persistence, but rather a more general
sense of agreement)'' \cite[\S 3.3]{RFC2418}.

The directives of the International Organization for Standardization (ISO)
and International Electrotechnical Commission (IEC) define consensus as
``General agreement, characterized by the absence of sustained opposition to
substantial issues by any important part of the concerned interests and by a
process that involves seeking to take into account the views of all parties
concerned and to reconcile any conflicting arguments.  NOTE Consensus need
not imply unanimity'' \cite[\S 2.5.6]{ISOIECDirectives}.

Andersson et al.\ did a case study of ISO information security standard
development from the perspective of the Swedish Institute for Standards
technical committee TC318 and found that, while consensus discourse
(``focusing on participation and consensus-making, such as `discuss' and
`harmonize'{}'') could be distinguished from warfare discourse (``military
metaphors such as `put pressure on,' `guard,' and `monitor'{}''), most
participants were silent and passive either way (cf.\ ``passive
non-resistance'' in \autoref{unanimity}) \cite{2020_Andersson}:

\begin{quote}
Choosing concepts such as resistance in a definition of consensus clearly
shows how the warfare discourse interacts with, and influences, the consensus
discourse.  This interaction between the two discourses also makes the
definition vague and ambiguous, leaving much room for interpretation; for
example: What is an absence of strong resistance? How strong is strong? What
are the important subject matters?  \dots\@  This example shows that even
though the consensus discourse was heavily emphasized and communicated to the
TC318 members, in practice, this was not the way de jure information security
standards were developed.  Instead, these best practices were developed by a
relatively small group of active members who took it upon themselves to
interpret when resistance to a proposal was too strong or was made by an
important stakeholder.
\end{quote}

Although the ``unity, not unanimity'' process of consensus is sometimes
attributed to the meeting practices of Quakers, a current Quaker book of
discipline states ``We are not seeking a consensus; we are seeking the will
of God'' \cite[\S 2.89]{QuakerBoD}.

\subsection{Contrast with gatekeeping}
\label{gatekeeping}

With consensus, a whole group speaks with one voice.  With gatekeeping,
dissenting voices are suppressed so that only one voice can be heard.

If the decision of a deliberative body is subject to arbitrary nullification,
the arbiter can keep rejecting decisions until the deliberative body delivers
what they want.  The deliberative body then serves no purpose but to launder
the will of the arbiter as their own putative consensus.

There can be no meaningful consensus when there is such an imbalance of power
that consent can be manufactured or compelled, where one party needs the
consent of the other, but the reverse isn't true.

These situations call into question the definition of consent itself, which
has been litigated to different conclusions in different jurisdictions.  For
example, in the United States, it is standard procedure for government
systems to present a dialog to government employees stating, ``You understand
and consent to the following:  \dots'' from which the only way to proceed is
to click a button labelled ``Accept.''  But in the European Union, consent
under such conditions is invalid \cite{EU_consent}:

\begin{quote}
  The employer-employee situation is generally considered as an imbalanced
  relationship in which the employer wields more power than the employee.
  Since consent has to be freely given, and in light of the imbalanced
  relationship, your employer in most cases can’t rely on your consent to use
  your data.
\end{quote}

\subsection{Qualitative methods of characterizing consensus}
\label{other}

Where a qualitative definition of consensus is used, someone is the final
arbiter of consensus.  It is a case of expert judgment substituting for
measurement.

IETF\@:  ``Consensus can be determined by a show of hands, humming, or any
other means on which the WG agrees (by rough consensus, of course).  Note
that 51\% of the working group does not qualify as `rough consensus' and 99\%
is better than rough.  It is up to the Chair to determine if rough consensus
has been reached'' \cite[\S 3.3]{RFC2418}.  RFC 7282, which is informational
and does not overrule the preceding text from RFC 2418, argues that the
determinant of rough consensus is the soundness of the technical case, not
the number of votes for or against \cite{RFC7282}.  ``One hundred people for
and five people against might not be rough consensus'' and ``Five people for
and one hundred people against might still be rough consensus'' are section
titles.

ISO/IEC\@:  ``If the leadership determines that there is a sustained
opposition, it is required to try and resolve it in good faith.  However, a
sustained opposition is not akin to a right to veto.  The obligation to
address the sustained oppositions does not imply an obligation to resolve
them successfully.  \dots\@ The responsibility for assessing whether or not a
consensus has been reached rests entirely with the leadership'' \cite[\S
  2.5.6]{ISOIECDirectives}.

Hand signals used in street protests lack standardization but may be more
expressive of individual sentiment than formal voting; e.g., an ordinal scale
consisting of \emph{agree}, \emph{don't agree}, \emph{oppose}, and
\emph{block} instead of simply yes or no \cite{Occupy}.

\subsection{Vulnerabilities}
\label{wrecking1}

Tactics used to wreck a consensus process include fraud, obstruction,
obfuscation, disinformation, flooding (exhausting the group with endless,
repetitive arguments), intimidation, and politicizing the question in terms
of a manufactured or exaggerated enemy.  See also the ``Ploys to subvert
consensus'' chapter of \cite{CWC}.

In groups using a qualitative definition of consensus, when a single member
chooses to obstruct the work of the group, the onus falls on the leadership
to resolve the problem.

IETF\@:  ``While open discussion and contribution is essential to working group
success, the Chair is responsible for ensuring forward progress.  When
acceptable to the WG, the Chair may call for restricted participation (but
not restricted attendance!)\ at IETF working group sessions for the purpose
of achieving progress.  The Working Group Chair then has the authority to
refuse to grant the floor to any individual who is unprepared or otherwise
covering inappropriate material, or who, in the opinion of the Chair is
disrupting the WG process.  The Chair should consult with the Area
Director(s) if the individual persists in disruptive behavior'' \cite[\S
  3.2]{RFC2418}.  Procedures for mailing list suspension were added in
\cite{RFC3934}.

\section{Quantifying consensus}
\label{newsec3}

\subsection{Definitions}

\subsubsection{General}

Quantitative definitions of consensus are those that are realizable in an
algorithmic sense.  They depend only on counts, amounts, and proportions, not
on expert judgment or the mood of a crowd.

Voting can be thought of not merely as a process for decision-making but as
the measurement method for a quantity called consensus.  From this viewpoint,
the vote is a dimensionless unit of measurement.  Once votes for each choice
have been counted, the result of the measurement can be expressed in various
ways:

\begin{ul}
\item As the number of votes for each choice (a count);
\item As the proportion of the total number of votes that each choice received \emph{or} the proportion of voters who voted for each choice (ratios);
\item As an ordinal ranking of choices by the number of votes received;
\item As a reduction to a dichotomic scale (e.g., a [sufficient] consensus
  does or does not exist).  Any ratio measurement can be reduced to a
  dichotomic scale by applying a threshold.
\end{ul}

None of these results is inextricably bound to the making of a particular
decision.  The results of measurements, including measurements of consensus,
are used to drive decisions, but the decision rules are separate from the
measurement.  Measurements provide data that can be used to inform decisions
or upon which a data-driven policy can be implemented.  A decision might be
required regardless of whether a consensus exists or is measurable.

That being said, the difference between the following two descriptions
usually reduces to semantics:

\begin{enumerate}[nosep,itemsep=6pt,leftmargin=*]
  \item Consensus is measured as a count, amount, or proportion.  Because
    this quantity exceeds a threshold, the decision is made that a motion
    passes.  In pseudocode:

    \vspace{3pt}
    \hspace*{1em} \texttt{Number Consensus1 = Measurement1()\ \ \# Measurement}\newline
    \hspace*{1em} \texttt{if Consensus1 > Threshold\ \ \ \ \ \ \ \ \ \ \ \# Condition}\newline
    \hspace*{2em} \texttt{motion passed \hspace*{12.26em} \# Decision}

  \item Consensus is measured on a dichotomic scale (there is or is not a
    consensus).  Because there is a consensus, the decision is made that a
    motion passes.  In pseudocode:

    \vspace{3pt}
    \hspace*{1em} \texttt{Boolean Consensus2 = Measurement2()\ \ \# Measurement}\newline
    \hspace*{1em} \texttt{if Consensus2\ \ \ \ \ \ \ \ \ \ \ \ \ \ \ \ \ \ \ \ \ \ \ \ \# Condition}\newline
    \hspace*{2em} \texttt{motion passed \hspace*{12.9em} \# Decision}
\end{enumerate}

If the dichotomic Consensus2 in description 2 is reduced from the count,
amount, or proportion that was used in description 1—that is to say, if
Measurement2 has the following pseudocode definition—then the behavior is
unchanged, and the only difference is that the thresholding operation was
moved from the decision rule into the measurement:

\hspace*{1em} \texttt{Boolean function Measurement2 = (Measurement1() > Threshold)}

But description 2 does not exclude the possibility that a dichotomic result
could be arrived at through a significantly different process.

If the population is evenly divided between two candidates in an election for
public office or if no choice is supported by a majority, there is no
consensus by any definition.  Voting reveals the absence of consensus in
these cases.  But the decision rules given by election law in practice will
pick a winner arbitrarily—sometimes literally by coin toss—rather than leave
an office vacant.

\subsubsection{Unanimity}
\label{unanimity}

Unanimity is obviously measurable as it entails the consent of every member
of the population.  There is no wiggle room on the unanimous part; there
is only variation in what constitutes consent.

The Oxford English Dictionary defines consensus (the most relevant sense) as
``Agreement in opinion, feeling, or purpose among a group of people, esp.\ in
the context of decision-making.  Also:  the collective unanimous opinion of a
number of people'' \cite[defn.~1.a]{OED3-consensus}.

When consensus is defined as unanimity, a defender of the status quo has the
power of a dictator.  Opposition can be sustained indefinitely by any member
who simply refuses to stop objecting.  Reconciliation can be prevented
indefinitely by any member who simply refuses to accept any offered
compromise.  A single member who is willing to sustain objection can say ``my
way or the highway.''  The result is an imbalance of power in favor of vested
interests (preservation of the status quo).

The Group of 20 (G20) uses consensus = unanimity:  ``All votes regarding the
report will proceed as a consensus form which means, one `Against' vote can
make [does make] the report fail'' \cite{2018_G20}.

The European Parliament uses consensus = unanimity, but there is a movement
to lower the threshold for many actions \cite{PE659451}.

The World Intellectual Property Organization (WIPO) has been requiring
unanimity for accreditation of observers, but this is not traceable to its
documented rules of procedure \cite{WIPO_GROP,WikFound2022}.

In some collaborative and decision-making bodies, consensus means
\emph{unanimous consent}, which is a term of art for a particular procedural
shortcut in which any who are opposed to the question are given the
opportunity to let it pass without objection.  Under Robert's Rules
\cite{Roberts}, a motion can pass by unanimous consent with only passive
silence from the majority of members; in other words, silence implies
consent.

The circumstantial difference between explicitly voting for something and
merely remaining silent when unanimous consent is called for is one basis for
claims that consensus is different than unanimous support.  It boils down to
where on the scale between enthusiastic, active support and detached, passive
non-resistance the threshold of individual consent resides.  It also gives
rise to the Abilene paradox, wherein the ostensible consensus is not actually
supported by any member of the group \cite{Harvey1974}.

\subsubsection{Unanimity minus a constant}
\label{UnanMinus}

The variation in which there is a fixed threshold for the number of
dissenting votes that is deemed compatible with consensus is documented in
Wikipedia \cite{WikiDecisionRules}.  Although it is a logical mitigation to
the worst pathologies of requiring unanimity, we have not found a definitive
example of its use in practice.  Without citation, Wikipedia alleges its use
in larger Masonic lodges:  ``Whilst in many such cases even a single black
ball will be fatal to the candidate's election, rules in larger clubs ensure
that a single member cannot exercise a veto to the detriment of the future of
the club.  For example, two black balls are required to exclude\dots''
\cite{WikiBlackball}.

\subsubsection{Supermajority proportion}

Consensus might be defined as the agreement of a supermajority such as
$\frac{3}{5}$, $\frac{2}{3}$, or $\frac{3}{4}$ of the population.  Although
decisions are made and public offices filled based on $\frac{1}{2}$ of the
population or less, such proportions are not regarded as indicating
consensus.

IETF uses a definition of ``rough consensus'' that is mostly qualitative, but
there is a note that ``51\% of the working group does not qualify as `rough
consensus' and 99\% is better than rough'' \cite[\S 3.3]{RFC2418}.

An amendment to the U.S. constitution must be ratified by $\frac{3}{4}$ of
the states.

\subsection{Quantitative methods of characterizing consensus}
\label{quantmethods}

In the following subsections we focus on common voting methods that can be
adapted to the measurement point of view.

\subsubsection{Simple example}
\label{simplmodl}

This section describes a measurement of consensus on a yes-or-no question
using a minimum number of votes as the quorum and a supermajority of votes as
the threshold for a decision.  Since quorum is determined from the number of
votes, abstentions (present but not voting) have no impact in this example.

Assume that the following values are given:

\begin{tabular}{ll}
$V_\mathrm{Y}$ & Number of votes \emph{in favor of} the proposition (``yes votes'') \\
$V_\mathrm{N}$ & Number of votes \emph{against} the proposition (``no votes'') \\
$T$ & Consensus threshold, $\frac{1}{2} < T \leq 1$ \\
$Q$ & Number of votes required for quorum, $Q \geq 1$
\end{tabular}

The measure of consensus is the proportion of votes in favor, $p =
V_\mathrm{Y} / (V_\mathrm{Y} + V_\mathrm{N})$.  This measure is reduced to a
dichotomic scale (consensus does or does not exist) by the following decision
rules, which furthermore associate the consensus (if it exists) with a
particular outcome:

\begin{tabular}{ll}
Condition & Decision \\ \hline
$V_\mathrm{Y} + V_\mathrm{N} \geq Q$ and $p \geq T$ &
  A consensus exists \emph{in favor of} the proposition. \\
$V_\mathrm{Y} + V_\mathrm{N} \geq Q$ and $(1-T) < p < T$ &
  Negative result.  There is evidence of the absence of consensus. \\
$V_\mathrm{Y} + V_\mathrm{N} \geq Q$ and $p \leq (1-T)$ &
  A consensus exists \emph{in opposition to} the proposition. \\
$V_\mathrm{Y} + V_\mathrm{N} < Q$ &
  Null result.  There is an absence of evidence.
\end{tabular}

\subsubsection{Elaborated model for yes-or-no questions}
\label{yn2}

To handle other methods of voting on yes-or-no questions, we introduce
another parameter, $P$, the effective population size.  It can be set at one
of four levels:  P(1), the nominal size of the voting body; P(2), the current
size of the voting body with vacant positions excluded; P(3), the number of
members present at the time of voting; or P(4), the number of members that
did not abstain.

A quorum can be defined as either (1) a minimum number of members that must
be present at the time of voting or (2) a minimum number of members that must
not abstain.  However, in any event, at least one member must vote for there
to be a quorum:  complete absence of evidence supports only a null result.

If the voting body does not have defined membership, P(1) and P(2) are not
applicable.  If abstentions are not counted, P(3) and quorum type (1) are not
applicable.

Quorum type (1) can be specified as a constant or it can be derived as a
proportion of P(1) or P(2).  In contrast, quorum type (2) can be specified as
a constant or it can be derived as a proportion of P(1), P(2), or P(3).
\autoref{Qtab} shows why certain combinations of P's and Q's are not valid.
Note that the effective population size used for quorum purposes need not be
the same as is used for determining consensus.

To cover all cases, we reformulate the model in terms of three logical
propositions:

\begin{tabular}{ll}
$q$           & Quorum is met \\
$t_\mathrm{a}$ & Threshold of consensus is met for acceptance \\
$t_\mathrm{r}$ & Threshold of consensus is met for rejection
\end{tabular}

Retaining these definitions from \autoref{simplmodl}:

\begin{tabular}{ll}
$V_\mathrm{Y}$ & Number of votes \emph{in favor of} the proposition (``yes votes'') \\
$V_\mathrm{N}$ & Number of votes \emph{against} the proposition (``no votes'')
\end{tabular}

The generalized threshold expressed in terms of $P$ (however $P$ is defined
for consensus purposes) can be a simple majority, a supermajority,
near-unanimity (unanimity minus a constant, see \autoref{UnanMinus}), or
unanimity.  If supermajority is used, we need:

\begin{tabular}{ll}
$T$ & Threshold for a supermajority as a proportion of $P$, $\frac{1}{2} < T \leq 1$
\end{tabular}

But if near-unanimity is used, we instead need:

\begin{tabular}{ll}
$C$ & Tolerated shortfall in votes from unanimity, $0 \leq C < \frac{1}{2}P$
\end{tabular}

The value of $T$ or $C$ would typically be fixed by voting rules while the
value of $P$ may vary.  The situation where $P$ becomes so small that the
constraint $C < \frac{1}{2}P$ is violated must be avoided through appropriate
choices of $P$ and $Q$ or by reducing $C$ in this special case.

Unanimity is the limiting case for both supermajority with $T=1$ and
near-unanimity with $C=0$:

\begin{tabular}{lll}
  Threshold      & $t_\mathrm{a}$ criterion      & $t_\mathrm{r}$ criterion      \\ \hline
  Majority       & $V_\mathrm{Y} > \frac{1}{2}P$ & $V_\mathrm{N} > \frac{1}{2}P$ \\
  Supermajority  & $V_\mathrm{Y} \geq TP$        & $V_\mathrm{N} \geq TP$        \\
  Near-unanimity & $V_\mathrm{Y} \geq P-C$       & $V_\mathrm{N} \geq P-C$       \\
  Unanimity      & $V_\mathrm{Y} = P$            & $V_\mathrm{N} = P$
\end{tabular}

The decision rules then become (with $\wedge$, $\vee$, and $\neg$ indicating
logical conjunction, disjunction, and negation, respectively):

\begin{tabular}{ll}
Condition & Decision \\ \hline
$q \wedge t_\mathrm{a}$ & A consensus exists \emph{in favor of} the proposition. \\
$q \wedge \neg (t_\mathrm{a} \vee t_\mathrm{r})$ & Negative result.  There is evidence of the absence of consensus. \\
$q \wedge t_\mathrm{r}$ & A consensus exists \emph{in opposition to} the proposition. \\
$\neg q$ & Null result.  There is an absence of evidence.
\end{tabular}

If P(1), P(2), or P(3) is used as the effective population size for
determining consensus, abstaining has the same impact as voting no.  But if
P(4) is used, only no-votes count against consensus.

\begin{table}
\begin{center}
\caption{Interactions between definitions of effective population size and quorum.}
\label{Qtab}

\vspace{3mm}
\begin{tabularx}{\textwidth}{|G|G|G|} \hline
  Specification of quorum &
  Quorum type (1), a minimum number of members that must be present at the time
  of voting &
  Quorum type (2), a minimum number of members that must not abstain \\ \hline\hline
  
  Constant & Valid & Valid \\ \hline
  
  Proportion of P(1), the nominal size of the voting body & Valid & Valid \\ \hline
  
  Proportion of P(2), the current size of the voting body with vacant positions
  excluded & Valid & Valid \\ \hline
  
  Proportion of P(3), the number of members present at the time of voting &
  Invalid:  proportion of members present that must be present &
  Valid \\ \hline
  
  Proportion of P(4), the number of members that did not abstain &
  Invalid:  proportion of members that did not abstain that must be present &
  Invalid:  proportion of members that did not abstain that must not abstain \\ \hline
\end{tabularx}
\end{center}
\end{table}

\subsubsection{Extended to 1-of-M contests}
\label{1ofM}

A yes-or-no question is a contest between two exclusive choices.  We now
consider the case where there are two or more choices, but only one can be
chosen.  In some voting terminology these are called 1-of-M contests
\cite{2021_EAC}.

However many choices the contest has, it is possible for the choice that
receives the most votes to satisfy one of the thresholds listed in
\autoref{yn2}.  We need only adjust the model to allow for an arbitrary
number of choices.  On the other hand, it is impossible to conclude from
votes in favor of various choices that they were voting \emph{against}
the others, so the criteria of rejection for yes-or-no questions do not
generalize to 1-of-M contests.

The logical propositions:

\begin{tabular}{ll}
$q$   & Quorum is met \\
$t_i$ & Threshold of consensus is met for choice $i$ \\
\end{tabular}

The measurables:

\begin{tabular}{ll}
$V_i$ & Number of votes for choice $i$ \\
\end{tabular}

The thresholds:

\begin{tabular}{ll}
  Threshold      & $t_i$ criterion \\ \hline
  Majority       & $V_i > \frac{1}{2}P$ \\
  Supermajority  & $V_i \geq TP$ \\
  Near-unanimity & $V_i \geq P-C$ \\
  Unanimity      & $V_i = P$
\end{tabular}

The decision rules:

\begin{tabular}{ll}
Condition & Decision \\ \hline
$q \wedge t_i$ & A consensus exists \emph{in favor of} choice $i$. \\
$q \wedge \neg (\bigvee t_x)$ & Negative result.  There is evidence of the absence of consensus. \\
$\neg q$ & Null result.  There is an absence of evidence.
\end{tabular}

When $M = 2$, the logic is similar to that of a yes-or-no question, but the
format is different.  The choices usually are candidates for office.

\subsubsection{N-of-M contests}

We now consider the case where there are two or more choices and two or more
can be chosen.  In some voting terminology these are called N-of-M contests
\cite{2021_EAC}.  The ballot instructions would typically say ``Vote for at
most $N$'' substituting the specific number for $N$.

In public elections that require only a plurality, it is trivial to change
the decision rules to fill $N$ offices.  One merely moves the line downward
to elect the top $N$ candidates:

\begin{tabular}{lr}
  Candidate & Votes \\ \hline\hline
  Montego Placeholder & 90\,785 \\
  Random Incumbent & 69\,212 \\
  Emilia Somebody & 49\,086 \\ \hline
  Ace Fakename & 46\,995 \\
  Known Character & 28\,479 \\
  Perennial Favorite & 5\,662
\end{tabular}

Applying the criteria given in \autoref{1ofM}, all, one, or none of the
elected candidates might satisfy a consensus threshold by themselves.  But
looking at the slate that was elected as a group, it is not necessarily the
case that any cast ballot included votes for that exact slate, that a
supermajority would prefer it to any other combination, or that the slate
elected is a rational choice for a cohesive team.  So, it is easy to identify
``consensus candidates'' from the individual tallies, but to identify a
``consensus slate'' if one exists requires tabulation by slate (the power set
of the set of candidates) rather than by individual candidate.

\subsubsection{Ranked order voting}
\label{RankedOrder}

Some public elections have voters rank choices in order of preference.  In
these cases, the vote as a unit does not arise until the tallying stage, at
which point the process begins by counting the number of ballots that ranked
each choice first.  As the process continues, if a choice is eliminated, the
votes for that choice are effectively transferred to the next choice in order
of preference for each ballot.

If ranked order voting is used, the criteria given in \autoref{1ofM} can be
applied to the counts derived from the most preferred choice of every ballot
in the first round of tabulation.  But if a winner is not selected in the
first round, it is fair to say that there is no consensus.  The ultimate
selection will be an explainable compromise but not a consensus choice.

\subsection{Limits of validity}

\subsubsection{Representation of voter intent}
\label{voterintent}

An accurately recorded vote tells you that someone voted that way.  It does
not tell you why they did.  There are many additional levels of validity that
one can ask for; e.g.:

\begin{ul}
\item The vote must be \emph{intentional}, i.e., not an accidental miscasting
  caused by misunderstanding the ballot or pressing the wrong button.

\item The vote must be \emph{informed}, i.e., not swayed by disinformation or
  ignorance.

\item The vote must be \emph{voluntary}, i.e., not swayed by coercion,
  pressure, or meddling of any kind.

\item The vote must be \emph{responsible}, i.e., not an arbitrary or
  capricious choice.

\item The vote must be \emph{intelligent}, i.e., not based on faulty reasoning.

\item The vote must be \emph{sincere}, i.e., not a ``strategic vote.''
\end{ul}

No democratic voting process can ensure a valid result when over half of
voters are casting votes that are compromised in one of the aforementioned
senses.

These additional conditions fundamentally change the problem from a
classical, physical measurement to one that is better approached with the
methods of social science.  There is a great deal of experience that can be
applied if the time and resources are available.  In practice, the applicable
laws, rules, and procedures determine how (or whether) these additional
conditions are addressed, and the variability is enormous.

\subsubsection{Impossibility theorems}

A series of formal results
\cite{Arrow1950,1970_Sen,1973_Gibbard,1975_Satterthwaite,2000_Duggan} has
shown that no voting method within broad classes of voting methods can
satisfy every criterion that one would want.  This means that every member of
some set of pathologies cannot simultaneously be prevented.

Undesirable properties that voting methods can have include:

\begin{ul}
\item The result can be determined by a single voter (a dictator) regardless
  of how others vote
\item The result can be option $Y$ even though every voter prefers $X$ to $Y$
\item A decision between $X$ and $Y$ can be influenced by factors other than
  the voters' rankings of $X$ and $Y$ relative to each other
\item The most effective way of voting to obtain a desired result is to vote
  insincerely for something else (``strategic voting'')
\end{ul}

Every voting method exhibits some pathology when the best result is not
obvious.  Fortunately, the goal of measuring consensus is not to decide a
winner in every circumstance but to determine whether circumstances indicate
a clear winner.

\subsubsection{Danger of surrogate measurement}
\label{surrogate}

The existence of both quantitative and qualitative definitions of consensus
suggests that the quantities could be regarded as \emph{surrogate measures}
for the unmeasurable qualities.  The meaning and ramifications of surrogate
measures are explained by the following two quotations, which were found via
\cite{Kaner}:

\begin{quote}
  Many of the attributes we wish to study do not have generally agreed
  methods of measurement.  To overcome the lack of a measure for an
  attribute, some factor which can be measured is used instead.  This
  alternate measure is presumed to be related to the actual attribute with
  which the study is concerned.  These alternate measures are called
  surrogate measures. \cite{Johnson}
\end{quote}

\begin{quote}
  It is usual that once a scale of measurement is established for a quality,
  the concept of the quality is altered to coincide with the scale of
  measurement.  The danger is that the adoption in science of a well defined
  and restricted meaning for a quality\dots\ may deprive us of useful insight
  which the common natural language use of the word gives us.
  \cite{Finkelstein}
\end{quote}

When surrogacy goes awry, we are left with an instance of the streetlight
effect, also known as the lamppost problem:  we are measuring a given
quantity not because it is what we were looking for but because it is what
we know how to quantify (it is where the light is).

\section{Measurement}
\label{newsec4}

\subsection{Scenarios}

Consensus is needed in different situations and is transformed by the
information available, the constraints on measurement, and the possible
outcomes of those situations.  The most relevant scenarios are described in
the following subsections.

\subsubsection{Public elections}

Public elections are a form of social choice with the following distinct
characteristics:

\begin{ul}
\item The voting population is large enough to make the measurement nontrivial
\item There is no quorum
\item Ballot secrecy complicates validation and verification
\item Votes are held infrequently and at significant cost
\item Contests are to choose one or more candidates from a slate of candidates,
  optionally followed by yes-or-no ballot questions
\item The contests, questions, and election process are fixed in advance through a legal process
\item Certain tactics for subverting a valid consensus, such as
  disinformation and politicizing a question (see \autoref{wrecking1}), are
  tolerated
\end{ul}

\subsubsection{Agreement of a committee or group}
\label{groupagr}

Agreement of a committee or group, including both organized meetings with
defined membership and ad hoc gatherings without defined membership, is a
form of social choice with the following distinct characteristics:

\begin{ul}
\item The voting population is small enough to eliminate many sources of uncertainty
\item A particular quorum is usually required
\item Ballot secrecy is rarely required
\item Votes can be held more frequently with less overhead
\item Most contests are yes-or-no questions
\item The questions are put forward by the same group and can be amended
\item The population has more power to change the process
\item Wrecking tactics may be subject to disciplinary action
\end{ul}

\subsubsection{Polls on social media}

Social media platforms including Mastodon, X, and Facebook\footnote{
Commercial entities or products are identified in order to describe a
measurement concept adequately.  Such identification does not imply
recommendation or endorsement by the National Institute of Standards and
Technology, nor does it imply that the entities or products are necessarily
the best available for the purpose.
} support polls with the following distinct characteristics:

\begin{ul}
\item Except for polls in closed groups, the size of the voting population is
  unpredictable—polls can go viral
\item There is no quorum
\item Ballot secrecy is optional and varies by platform
\item Polls can be created at will by anyone at virtually no cost
\item Contest types are limited by the platform
\item The barrier to voting multiple times using different accounts is low to
  nonexistent, so the process is not appropriate for high-stakes decisions
\end{ul}

For example, in Mastodon version 4.2.10, polls have the following limitations:

\begin{ul}
\item The supported contest types are 1-of-M and M-of-M (i.e., N-of-M where N
  = M; all choices may be selected simultaneously)
\item The number of choices is limited to 4 (configurable in the \href{https://glitch-soc.github.io/docs/}{glitch-soc fork})
\item Closed groups are not yet supported except by controlling access to the whole Mastodon server
\item No write-ins—additional options are customarily suggested in replies but
  are not measured as part of the poll
\item No counting of abstentions
\end{ul}

Closed groups, abstentions, and additional contest types have been requested
(see issues
\href{https://github.com/mastodon/mastodon/issues/19059}{\#19059},
\href{https://github.com/mastodon/mastodon/issues/19952}{\#19952},
\href{https://github.com/mastodon/mastodon/issues/20519}{\#20519},
\href{https://github.com/mastodon/mastodon/issues/23295}{\#23295}, and
\href{https://github.com/mastodon/mastodon/issues/25339}{\#25339}), but their
implementation will or would require changes to the many client apps as well
as to the database schema and server.

\subsection{Preconditions and process control}
\label{precon}

The context and procedures that apply to a measurement of consensus can
affect the validity of the measurement both in terms of the values obtained
and in terms of their fitness for purpose.  As with any measurement, it is
necessary for confounding factors to be under control.

\subsubsection{Ensuring fair ballot access}

Ballot access refers to the ability for all relevant options to be presented
to voters.  In public elections, there are criteria and procedures in
election law that limit which candidates or issues are listed on the ballot.
In committee or group votes, the applicable rules limit which questions can
get on the agenda and be brought up for vote.

We can identify two different levels at which consensus can operate.  The
first is the level of an individual question that is put to a vote.  The
second is the level of the overall business or goals of the group or
population that is voting.  Accurate measurement of consensus at the first
level is necessary but insufficient.  If the options that are actually most
preferred are not on the ballot, you can measure consensus on the question
presented, but it is still the wrong question.

A consensus that $X$ is better than $Y$ does not imply that there is a
consensus in favor of $X$.  $X$ might have lost to a third option $Z$ or to
``none of the above'' if these options had been offered.  Similarly, a lack
of support for any choice offered in the slate of options for moving forward
does not imply that there is a consensus in favor of maintaining the status
quo.

Given control of the ballot and some not very restrictive conditions, the
McKelvey–Schofield theorem shows that choices made by majority vote can be
forced arbitrarily far away from the preferences of a population by playing
subpopulations against one another in successive polls
\cite{1976_McKelvey,1978_Schofield}.

Biased control of the agenda and the ballot is thus incompatible with valid
measurement of consensus.

\subsubsection{Mitigating obstruction}
\label{wrecking2}

If unanimous support is not a prerequisite for consensus, it is necessary to
define procedurally how a single member who chooses to obstruct the work of a
group can be overruled in a fair and orderly fashion.  This is
straightforward when consensus is measured through voting—just use a
threshold that is less than unanimity.  However, obstruction can occur
through means other than the exercise of an individual vote, e.g.,
quorum-busting \cite{WikiQuorumBusting,walkout}, which may require more
extensive procedural mitigations.

\subsubsection{Avoiding sequential voting pathologies}

In organized meetings, when a choice must be made among exclusive options,
house rules often force the contest to be broken into a sequence of yes-or-no
votes on the individual options.  This results in a number of serious
pathologies, which are described briefly in the following paragraphs.  When
options are exclusive, it is better to design a single contest that allows a
direct choice among them.

\textbf{Conflicting choices.}  A consensus can exist in support of two or
more exclusive options.  If consensus was measured on a dichotomic scale, we
can go no farther than demonstrating that this situation exists.  If ordinal
or ratio scales were used, we can compare the results to possibly determine
which consensus is stronger.

\textbf{Suboptimal choice.}  If the decision is made as soon as any of the
options receives a passing result, the process is biased in favor of
whichever options are offered earlier in the order of voting.  This follows
directly from the possibility of two or more of the options each having
adequate support.

If the full list of options to be considered is not made available to the
group before the start of sequential voting, we have an instance of the
secretary problem \cite{SecretaryProblem}.  At each step, the group is faced
with the dilemma of settling for the choice before them or rejecting it in
hopes that something better will be offered.  If there is one option that
would unambiguously be the group's favorite (a strong Condorcet winner), the
probability of it being chosen by this process is less than half.

\textbf{Default.}  The result of sequential voting can be the group rejecting
every option and defaulting on its obligation to choose among them.  When all
of the options are unpopular, despite there being no alternatives or better
ideas on offer, sequential voting leads to paralysis.

\subsection{Example implementation of decision rules}
\label{implementation}

The decision rules for consensus with the contest types described in
\autoref{quantmethods} have been implemented in a Ruby module named
ConsensusMeasurement.  It includes only the logic for evaluating consensus
given the parameters discussed in \autoref{quantmethods}, not the collection
and tabulation of ballots and votes.

The module and its associated test cases are available as a separate
distribution \cite{ConsensusGithub}.  The following excerpted comments and
method signatures describe the implementation and the adaptations made.

The simple model of \autoref{simplmodl} is implemented in a single method:

\begin{verbatim}
# Measurement of consensus on a yes-or-no question using a supermajority of
# votes as the threshold
# quorum               Integer >= 1
# votes_y and votes_n  Integer >= 0
# threshold            Rational 1/2 < T <= 1
# Returns :negative_result, :null_result, :accepted, or :rejected.
def self.question_simple(quorum, votes_y, votes_n, threshold)
\end{verbatim}

  The methods for the elaborated model don't take the nominal or current size
  of the voting body as parameters.  If the quorum is specified as a
  proportion of P(1) or P(2), the caller must reduce it to the number of
  members and specify quorum\_type :num\_present or :num\_voting as
  applicable.  For example, if quorum is defined as $\frac{1}{3}$ of P,
  Rational(P,3).ceil should be passed as the value for quorum.

  The remaining option is quorum type (2) defined as a proportion of P(3).
  This is supported by specifying quorum\_type :proportion\_voting and passing
  the proportion as a Rational value for quorum.

  Thus, the parameters related to the determination of quorum are:

\begin{tabular}{ll}
quorum\_type & :num\_present, :num\_voting, or :proportion\_voting \\
quorum       & Integer $\geq$ 1 (:num\_present and :num\_voting) \\
             & Rational, 0 $<$ proportion $\leq$ 1 (:proportion\_voting) \\
present and voting & Integer $\geq$ 0, present $\geq$ voting
\end{tabular}

  All options for determining consensus are supported by the following
  parameters:

\begin{tabular}{ll}
population & Integer P $\geq$ 0 \\
threshold\_type & :majority, :supermajority, :near\_unanimity, or :unanimity \\
threshold & For :near\_unanimity, Integer 0 $\leq$ C $<$ P/2 \\
          & For :supermajority, Rational $\frac{1}{2}$ $<$ T $\leq$ 1 \\
          & Optional and ignored for :majority and :unanimity
\end{tabular}

\begin{verbatim}
# Determine whether quorum is met
# Parameters as described in the notes above
# Returns boolean
def self.quorate?(quorum_type, quorum, present, voting)

# Evaluate threshold of consensus for a particular choice
# votes  Integer >= 0
# Other parameters as described in the notes above
# Returns boolean
def self.consensus?(votes, population, threshold_type, threshold=nil)

# Elaborated yes-or-no question
# votes_y and votes_n  Integer >= 0
# Other parameters as described in the notes above
# Returns :negative_result, :null_result, :accepted, or :rejected
def self.question(quorum_type, quorum, present, votes_y, votes_n, population,
                  threshold_type, threshold=nil)

# N-of-M contest
# (For 1-of-M, see one_of_m below)
# votes  Array of Integer >= 0 totalling at most N × voting
# Other parameters as described in the notes above
#
# Returns :negative_result, :null_result, or an array of indices of the
# choices in array votes that passed the threshold of consensus by
# themselves.  No attempt is made to identify a consensus slate if one
# exists; only the individual choices are evaluated.  
def self.n_of_m(quorum_type, quorum, present, voting, votes, population,
                threshold_type, threshold=nil)

# 1-of-M contest
# Mostly the same as n_of_m, but with the following simplifications for N=1:
# - Parameter voting is removed since it must equal votes.sum
# - Returns :negative_result, :null_result, or the index of the choice in
#   array votes that passed the threshold of consensus
def self.one_of_m(quorum_type, quorum, present, votes, population,
                  threshold_type, threshold=nil)
\end{verbatim}

In addition to basic type and range checks on inputs, various additional
constraints are enforced where feasible given the data provided to the
different methods:

quorate?\\
– Number voting must not exceed number present

consensus?\\
– Number of votes must not exceed size of population

question\\
– Number of votes must not exceed number present\\
– Number of votes must not exceed size of population

n\_of\_m\\
– Number voting must not exceed number present\\
– Number voting must not exceed size of population\\
– Number voting must not exceed number of votes\\
– Number of votes must not exceed M times number voting*

one\_of\_m invokes n\_of\_m with number voting = number of votes.

* Since N is not a parameter, the burden is partly on the caller to ensure
that the number of votes represented in the array votes is no greater than N
times the number voting.  The function consensus?\ will throw an exception if
any element of the array exceeds population.  n\_of\_m checks the upper bound
N = M.

The value of present (used for quorum) can exceed the value of population
(used for consensus) when P(4) the number of members that did not abstain is
used as the effective population size for determination of consensus.

\section{Uncertainty}
\label{newsec5}

\subsection{General}

Uncertainty is part of all elections and consensus processes.  As Bunnett
wrote in the aftermath of the 2000 U.S. presidential election, ``The
uncertainty of measurement is much greater than one vote. \dots\@ Scientific
realities are widely recognized in other aspects of our lives, so why not in
something so fundamental to the United States as the election
process?''~\cite{Bunnett2020}

One might set a goal of producing vote counts \emph{with uncertainty
intervals} for particular choices.  Given various inputs both actual and
assumed, one could calculate these.  But the subsequent application of
decision rules to create sharp boundaries militates against a rational
treatment of uncertainty.  To say that a decision is mandatory is to concede
that you will accept an arbitrary decision.  Even if there were no
uncertainty about the counts, a tied result would not support the making of a
decision.  Uncertainty intervals increase the number of situations in which
we consider counts to be effectively equal, where there is little or no
confidence in a finding that one is greater than the other.

Several different sources of uncertainty may apply:

\begin{ulv}
  \item When the number of votes is less than the population size, whether
    due to abstentions or to limited enfranchisement, there is sampling
    uncertainty in treating the consensus of voters as the consensus of the
    population.  It is not a given that the sense of the non-voting
    subpopulation is relevant to the question, as the next section will
    further explain, but one cannot legitimately claim to have evidence of
    consensus if the poll was widely boycotted.

  \item There is the possibility that the votes cast by individuals were not
    what they intended or would have intended in the absence of confounding
    factors.

  \item A voter might cast a vote that is unclear, creating uncertainty about
    the voter's intent.

  \item There is the possibility of individual votes being recorded,
    transmitted, or counted incorrectly.

  \item There is the possibility of fraud within the voting process.
\end{ulv}

\subsection{Treatment of abstentions}
\label{Unc1}

From a measurement viewpoint, yes means yes, no means no, and
\emph{null means null}.  We may have insufficient evidence to say whether or
not a consensus exists.  Uncertainty may completely occlude the result.

Interpreting silence as support of or opposition to a choice adds a positive
or negative bias to the outcome, respectively.  Typically, when a yes-or-no
question is being voted, the default choice is \emph{no} to a proposed
change.  Absence of support for a change does not imply consensus in favor of
the status quo, but the status quo is reinforced all the same, producing an
imbalance of power in favor of vested interests.

Compulsory voting may reduce abstention but does not necessarily ensure a
more valid result, as other properties enumerated in \autoref{voterintent}
will be undermined.

\subsection{In public elections}

The distinct characteristics of public elections create additional issues for
the management of uncertainty, but they also have a regulated process that
adds mitigations for uncertainty:

\begin{ulv}
\item The population that is allowed to vote is determined through voter
  registration.

\item Ballot instructions are provided according to best practices and
  regulations.  Beyond that, however, the burden is on voters to understand
  the ballot and to verify that their votes are correctly presented to the
  recording system.  Since votes are secret, there is no verification that
  the vote was cast as intended.

\item Public elections are highly scrutinized both by political parties and
  by election authorities.  The risk of errors in transmission or counting
  (such as those from poorly calibrated equipment, misprinting of ballots, or
  other systematic issues) is mitigated by audits performed before the vote
  is finalized.  There is, by design, no traceability from the published
  totals back to an individual's cast vote.  Cryptographic methods have been
  proposed to establish this traceability while preserving properties of
  privacy and secrecy (end-to-end auditable voting systems), but they are not
  standardized, and current methods provide only the ability for individual
  voters to check whether their votes were correctly counted.

\item Fraud is controlled through procedures and law enforcement.  Although
  some forms of fraud can be detected and their magnitudes estimated through
  statistical methods, it is typically reduced to a yes-or-no finding that an
  election is ``free and fair'' or not, separate from the results being
  reported.  Factors outside of the voting system such as coercion may be
  harder to detect.
\end{ulv}

Calculation of uncertainty intervals for counts in public elections is more
complicated than it at first appears:

\begin{ulv}
  \item False detection of a mark on a paper ballot can cause an overvote
    condition, which causes any valid vote in that contest to be discarded.
  \item Failure to detect a mark on a paper ballot can lead to an undervote
    condition or the counting of a vote when it was actually overvoted.
  \item If write-ins are allowed, the number of choices may not be fixed
    beforehand, and there are additional exceptions to deal with.
\end{ulv}

The acceptable error in a public election vote count, as related by some
election administrators, is one fewer than the margin of victory.  The closer
the election results, the greater the risk for errors to change the outcome.
Risk-limiting audits are a new mitigation that is required by new standards.
They limit the risk of an incorrect \emph{decision} \cite{Lindeman2012}.

\subsection{In committees and groups}

The distinct characteristics of voting in committees and groups work to
control some sources of uncertainty:

\begin{ulv}
\item In organized meetings, sampling uncertainty may be limited by a
  sufficiently large quorum.
\item Members can raise points of order and request clarification of any
  unclear aspects of the poll.
\item Since the process is conducted without secrecy, voters can in most
  cases witness that all votes were correctly recorded and counted.
\end{ulv}

\subsection{In social media polls}

In contrast, most sources of uncertainty are intensified in social media
polls:

\begin{ulv}
\item The size of the population and, hence, sampling uncertainty are
  unpredictable.

\item Ballot instructions are neither regulated by election law nor subject
  to clarification through procedural motions, so voters could easily be
  confused about what they are voting on or the impact of a choice.

\item A fault in the platform could produce systematic error in the
  recording, transmission, or counting of votes that, in the absence of any
  audit, would not be detected.

\item The barriers to fraud are low.
\end{ulv}

\section{Other uses of consensus}
\label{newother}

\subsection{Scientific consensus}

Concepts such as \emph{scientific consensus} or \emph{medical consensus}
pertain to what is generally accepted by the relevant population of
scientists or other professionals.  Such acceptance is practically never
unanimous.  Some level of skepticism of and divergence from widely-held
beliefs is considered healthy as long as it is consistent with available
empirical evidence.

In practice, scientific consensus is either characterized through the
agreement of a committee or group (\autoref{groupagr}) or asserted by
gatekeeper fiat (\autoref{gatekeeping}).  There is no attempt at measuring
the consensus of the population as a whole except inasmuch as a committee or
group might constitute a representative sample.

Measuring the weight of evidence behind a theory is independent of measuring
whether scientific consensus is with or against it.  Unfortunately, in
gatekeeping processes it can happen that an existing scientific consensus is
used as a measure of the plausibility of new evidence, with the result that a
self-sustaining orthodoxy develops that is resistant to or incapable of
change.

\subsection{Consensus values in international metrology}

The term \emph{consensus value} was introduced in international metrology to
refer to a statistically-derived estimate of a measurand that national
metrology institutes could agree to use \cite{Paule1982}:

\begin{quote}
The purpose of this article is to discuss the problem of calculating ``best''
estimates from a series of experimental results.  It will be convenient to
refer to these estimates as consensus values.
\end{quote}

Calculating such a value is a different thing than measuring the social
consensus of a group and is, thus, out of scope for this discussion.

\subsection{Consensus in multi-agent systems}

In computer science, \emph{consensus protocols} are used to decide an outcome
in the presence of unreliable communications and unreliable or malicious
agents.  Although the word \emph{consensus} is the same, the nature of the
problem is different.  There exists a ground truth in light of which the
actions of individual agents and the outcome of the process are deemed
correct or incorrect.  This means that social concepts of consent and
fairness to all parties concerned are out of scope, and software consensus
protocols are thus out of scope for this discussion.  But one concept that
can be generalized to other contexts is the fault tolerance threshold:  no
consensus process can ensure a correct decision if too many agents behave
incorrectly.

\section{Conclusion}
\label{newsec6}

This article investigated the nature and limitations of the measurement model
of consensus.  Beginning with qualitative approaches to consensus, we saw
that ``unity'' is an attractive goal, but it is difficult to distinguish it
from the elusive state of unanimous agreement.  We then proceeded to examine
definitions, measurement, and uncertainty for consensus as a quantity, which
can be measured in an automatable process and reduced to a dichotomic scale
through decision rules.

For those seeking the social qualities of unity and cooperation, the
compromises made in the process of quantifying consensus may amount to
entirely missing the point, making the quantity an illegitimate
\emph{surrogate measure} (\autoref{surrogate}).  But once unity and
cooperation become unobtainable for any reason, the alternatives to
measurement amount to the end of functioning as a group, one way or another
(paralysis, war, schism, or dissolution).

We conclude with a few observations that are true under any reasonable
definition of consensus:

\begin{ulv}
  \item If no choice is supported by a simple majority, there is no consensus.

  \item If a choice is being made between exclusive alternatives, consensus
    cannot exist without a strong Condorcet winner—one option that would
    unambiguously be the group's favorite.

  \item Biased control of the agenda and the ballot is incompatible with
    valid measurement of consensus.

  \item A group decision-making system in which a single participant can
    dictate the outcome or prevent a conclusion from being reached is not a
    consensus system.
\end{ulv}

\section*{Acknowledgments}

Thanks to NIST colleagues Paul E. Black, Barbara Guttman, and Raghu N. Kacker
for constructive feedback.

\bibliographystyle{unsrturl}
\bibliography{Consensus}

\end{document}